\documentclass[11pt,english]{revtex4}
\usepackage{lmodern}

\usepackage[T1]{fontenc}
\usepackage[latin9]{inputenc}
\usepackage{babel}

\usepackage{amsmath}
\usepackage{amssymb}
\usepackage{esint}
\usepackage[unicode=true, pdfusetitle,
 bookmarks=true,bookmarksnumbered=false,bookmarksopen=false,
 breaklinks=false,pdfborder={0 0 1},backref=false,colorlinks=false]
 {hyperref}

\makeatletter
\@ifundefined{textcolor}{}
{%
 \definecolor{BLACK}{gray}{0}
 \definecolor{WHITE}{gray}{1}
 \definecolor{RED}{rgb}{1,0,0}
 \definecolor{GREEN}{rgb}{0,1,0}
 \definecolor{BLUE}{rgb}{0,0,1}
 \definecolor{CYAN}{cmyk}{1,0,0,0}
 \definecolor{MAGENTA}{cmyk}{0,1,0,0}
 \definecolor{YELLOW}{cmyk}{0,0,1,0}
 }

\usepackage{latexsym}\usepackage{bm}

\makeatother

\usepackage{babel}

\makeatother

\begin{document}

\title{CANONICAL STRUCTURE OF HIGHER DERIVATIVE GRAVITY IN 3D}

\author{\.{I}brahim Güllü }

\email{e075555@metu.edu.tr}

\affiliation{Department of Physics,\\
 Middle East Technical University, 06531, Ankara, Turkey}

\author{Tahsin Ça\u{g}r\i{} \c{S}i\c{s}man}

\email{sisman@metu.edu.tr}

\affiliation{Department of Physics,\\
 Middle East Technical University, 06531, Ankara, Turkey}

\author{Bayram Tekin}

\email{btekin@metu.edu.tr}

\affiliation{Department of Physics,\\
 Middle East Technical University, 06531, Ankara, Turkey}

\date{\today}
\begin{abstract}
We give an explicitly gauge-invariant canonical analysis of linearized
quadratic gravity theories in three dimensions for both flat and de
Sitter backgrounds. In flat backgrounds, we also study the effects
of the gravitational Chern-Simons term, include the sources, and compute
the weak field limit as well as scattering between spinning massive
particles. 
\end{abstract}


\maketitle

\section{Introduction}

Recently, Bergshoeff \emph{et al.} \cite{bht} found that, in three
dimensions, among the class of higher-derivative theories defined
by the Lagrangian $\kappa^{-1}R+\alpha R^{2}+\beta R_{\mu\nu}^{2}$,
a special case $8\alpha+3\beta=0$ and $\kappa^{-1}<0$ (let us call
it BHT gravity) and its parity-violating extension, with a gravitational
Chern-Simons term, have massive ghost-free spin-2 particles in their
free spectrum around both flat and (anti)-de Sitter {[}(a)dS{]} spacetimes.
Perhaps, the most interesting feature of the BHT model is that it
is the first and (apart from some bimetric theories) the only known
example of a (parity-invariant) theory that provides a nonlinear extension
to the Pauli-Fierz mass term for spin-2 particles. In addition, being
a three-dimensional theory, it is powercounting superrenormalizable
whose four-dimensional cousin is renormalizable \cite{stelle}. Therefore,
it is possible that the BHT model may turn out to be a perturbatively
well-defined quantum gravity in three dimensions. But of course, unitarity
of the model beyond tree level is yet to be checked.

Various aspects of the theory such as its ghost-freedom and tree level
unitarity \cite{bht,deser,nakasone,gullu} and Newtonian limits \cite{gullu}
have been explored. Also, classical solutions and related issues were
studied in \cite{bht,liu-sun,giribet,clement,troncoso,gurses}, and
supergravity extensions were given in \cite{andringa}.

In this paper, we give an explicitly gauge-invariant, detailed analysis
of the canonical structure of the generic quadratic models in $2+1$
dimensions for both flat and de Sitter (dS) backgrounds. In flat space,
we also include the gravitational Chern-Simons term in our analysis.
It is interesting to see how at the linearized level BHT theory is
singled out as a unique regular {}``harmonic oscillator'' (massive
free field), which avoids the infamous Ostragradskian instability
that ruins every higher-time derivative theory \cite{woodard}. {[}It
was claimed that adding interactions might yield stable higher-time
derivative theories \cite{smilga}.{]} All the other quadratic theories
are ghost-ridden higher-derivative Pais-Uhlenbeck \cite{pais} oscillators
at the linearized level. In addition, we discuss the Newtonian limits,
weak fields, and the tree level scattering of particles with mass
and spin in these models.

The layout of the paper is as follows: Section \ref{sec:Flat-space}
is devoted to flat spacetime analysis which includes the canonical
structure of both the parity-invariant and parity-violating quadratic
gravity, in addition to the effects of static sources and weak field
solutions with circular symmetry. In Section \ref{sec:dS_space},
canonical structure analysis is extended to de Sitter space. Some
of the computations are relegated to the Appendices. Tree level scattering
amplitude between spinning massive particles is given in Appendix
A. In Appendix B, generic quadratic action is written in terms of
two auxiliary fields. Finally, we list some results which may be helpful
in the analysis of field equations.

\section{Higher-derivative spin-2 in flat spacetime\label{sec:Flat-space}}

We start our analysis of the higher-derivative spin-2 fields in flat
space, which is considerably simpler than the de Sitter background,
which we deal with in the next section. The action \begin{equation}
I=\int d^{3}x\,\sqrt{-g}\left(\frac{1}{\kappa}R+\alpha R^{2}+\beta R_{\mu\nu}^{2}\right),\label{eq:Non-linear_action}\end{equation}
 gives the desired spin-2 model when expanded as $g_{\mu\nu}=\eta_{\mu\nu}+h_{\mu\nu}$,
where $\eta_{\mu\nu}$ is the usual flat spacetime metric with mostly
plus signature. {[}Actually, spin-2 here is a misnomer. It should
be symmetric rank-2 tensor, since without any constraints in addition
to spin-2, it has spin-1 and spin-0 components. But, in what follows,
we will call $h_{\mu\nu}$ a spin-2 field.{]} Below, we will also
add the parity-violating gravitational Chern-Simons term to this action.
In practice, to actually get the action for $h_{\mu\nu}$, it is somewhat
more convenient to linearize the full nonlinear field equations and
then integrate them (after carefully taking care of the overall sign,
which will be relevant for the discussion of ghosts). Then, the action
(\ref{eq:Non-linear_action}) up to boundary terms becomes \begin{equation}
I=-\frac{1}{2}\int d^{3}x\, h_{\mu\nu}\left[\frac{1}{\kappa}\mathcal{G}_{L}^{\mu\nu}+\left(2\alpha+\beta\right)\left(\eta^{\mu\nu}\Box-\partial^{\mu}\partial^{\nu}\right)R_{L}+\beta\Box\mathcal{G}_{L}^{\mu\nu}\right].\label{eq:flat_sec_ord_action}\end{equation}
 Here, the linearized Einstein and Ricci tensors, and curvature scalar
read \begin{gather}
\mathcal{G}_{L}^{\mu\nu}=R_{L}^{\mu\nu}-\frac{1}{2}\eta^{\mu\nu}R_{L},\qquad R_{L}=\partial_{\alpha}\partial_{\beta}h^{\alpha\beta}-\Box h,\nonumber \\
R_{L}^{\mu\nu}=\frac{1}{2}\left(\partial_{\sigma}\partial^{\mu}h^{\nu\sigma}+\partial_{\sigma}\partial^{\nu}h^{\mu\sigma}-\Box h^{\mu\nu}-\partial^{\mu}\partial^{\nu}h\right),\qquad h=\eta^{\mu\nu}h_{\mu\nu},\label{eq:flat_lin_tensors}\end{gather}
 where $\Box=\partial_{\mu}\partial^{\mu}=-\partial_{0}^{2}+\nabla^{2}$.
Raising and lowering operations are carried out with $\eta_{\mu\nu}$.
To explore the canonical structure and identify the free fields, $h_{\mu\nu}$
can be decomposed in terms of six \emph{a priori} free functions of
$\left(t,\vec{x}\right)$:\begin{align}
h_{ij} & \equiv\left(\delta_{ij}+\hat{\partial}_{i}\hat{\partial}_{j}\right)\phi-\hat{\partial}_{i}\hat{\partial}_{j}\chi+\left(\epsilon_{ik}\hat{\partial}_{k}\hat{\partial}_{j}+\epsilon_{jk}\hat{\partial}_{k}\hat{\partial}_{i}\right)\xi,\nonumber \\
h_{0i} & \equiv-\epsilon_{ij}\partial_{j}\eta+\partial_{i}N_{L},\qquad h_{00}\equiv N,\label{eq:Metric_decomposition}\end{align}
 where $\hat{\partial}_{i}\equiv\partial_{i}/\sqrt{-\nabla^{2}}$.
From these, one can compute $\mathcal{G}_{\mu\nu}^{L}$ in terms of
three functions: \[
\mathcal{G}_{00}^{L}=-\frac{1}{2}\nabla^{2}\phi,\qquad\mathcal{G}_{0i}^{L}=-\frac{1}{2}\left(\epsilon_{ik}\partial_{k}\sigma+\partial_{i}\dot{\phi}\right),\]
 \[
\mathcal{G}_{ij}^{L}=-\frac{1}{2}\left[\left(\delta_{ij}+\hat{\partial}_{i}\hat{\partial}_{j}\right)q-\hat{\partial}_{i}\hat{\partial}_{j}\ddot{\phi}-\left(\epsilon_{ik}\hat{\partial}_{k}\hat{\partial}_{j}+\epsilon_{jk}\hat{\partial}_{k}\hat{\partial}_{i}\right)\dot{\sigma}\right],\]
 where $\dot{\phi}=\partial\phi/\partial t$, etc. Here, $q$, $\sigma$,
and $\phi$ are invariant under gauge transformations $\delta_{\zeta}h_{\mu\nu}=\partial_{\mu}\zeta_{\nu}+\partial_{\nu}\zeta_{\mu}$
and are defined as\begin{equation}
q\equiv\nabla^{2}N-2\nabla^{2}\dot{N_{L}}+\ddot{\chi},\qquad\sigma\equiv\dot{\xi}-\nabla^{2}\eta.\label{eq:flat_gauge_inv_comb}\end{equation}
 Note also that $\phi$ is gauge invariant unlike the other components
of $h_{\mu\nu}$. Linearized scalar curvature is computed to be\[
R_{L}=q-\Box\phi.\]
 Therefore, as required by the Bianchi identity, $\partial_{\mu}\mathcal{G}_{L}^{\mu\nu}=0$,
the number of arbitrary functions reduces from six to three. One can
use either $\phi$, $\sigma$, $q$; or $\phi$, $\sigma$, $R_{L}$
combinations. The Einstein-Hilbert part of the action can be computed
as \[
I_{EH}=-\frac{1}{2\kappa}\int d^{3}x\, h_{\mu\nu}\mathcal{G}_{L}^{\mu\nu}=\frac{1}{2\kappa}\int d^{3}x\,\left(\phi q+\sigma^{2}\right),\]
 which clearly shows that there is no propagating degree of freedom
in the pure Einstein theory. To compute the quadratic part, its better
to use the self-adjointness of the involved operators to rewrite the
action as explicitly gauge invariant not just gauge invariant up to
a boundary term, which will simplify the computations in a great deal:\[
I_{2\alpha+\beta}=-\frac{2\alpha+\beta}{2}\int d^{3}x\, h_{\mu\nu}\left(\eta^{\mu\nu}\Box-\partial^{\mu}\partial^{\nu}\right)R_{L}=\frac{2\alpha+\beta}{2}\int d^{3}x\, R_{L}^{2},\]

\begin{align*}
I_{\beta} & =-\frac{\beta}{2}\int d^{3}x\, h_{\mu\nu}\Box\mathcal{G}_{L}^{\mu\nu}=-\frac{\beta}{2}\int d^{3}x\,\left(-2\mathcal{G}_{\mu\nu}^{L}\mathcal{G}_{L}^{\mu\nu}+\frac{1}{2}R_{L}^{2}\right)=\frac{\beta}{2}\int d^{3}x\,\left(q\Box\phi+\sigma\Box\sigma\right).\end{align*}
 In the $I_{\beta}$ action, the second equality follows after one
moves the $\Box$ term to $h_{\mu\nu}$, and then uses (\ref{eq:flat_lin_tensors})
and the Bianchi identity. Collecting all the terms, the total action
in terms of the gauge-invariant combinations is\begin{align}
I & =\frac{1}{2}\int d^{3}x\,\left[\frac{1}{\kappa}\phi q+\left(2\alpha+\beta\right)\left(q-\Box\phi\right)^{2}+\beta q\Box\phi\right]+\frac{\beta}{2}\int d^{3}x\,\left(\sigma\Box\sigma+\frac{1}{\kappa\beta}\sigma^{2}\right).\label{eq:flat_total_action}\end{align}
 $\sigma$ describes a single scalar field with mass $m_{g}^{2}\equiv-\frac{1}{\kappa\beta}$
which is nontachyonic for $\kappa\beta<0$ and a nonghost for $\beta>0$,
therefore $\kappa<0$. For the $\phi$ and $q$ part of the action,
the discussion bifurcates whether $2\alpha+\beta=0$, or not. Let
us first consider the $2\alpha+\beta\ne0$ case, for which the nondynamical
field $q$ can be eliminated, yielding the action\begin{equation}
I_{\phi}=\frac{1}{2}\int d^{3}x\,\left[\frac{\beta\left(8\alpha+3\beta\right)}{4\left(2\alpha+\beta\right)}\left(\Box\phi\right)^{2}+\frac{\left(4\alpha+\beta\right)}{2\kappa\left(2\alpha+\beta\right)}\phi\Box\phi-\frac{1}{4\kappa^{2}\left(2\alpha+\beta\right)}\phi^{2}\right].\label{eq:phi_action}\end{equation}
 There are apparently several special points one of which is the BHT
limit $8\alpha+3\beta=0$, for which the higher-derivative term disappears.
{[}The $4\alpha+\beta=0$ theory seems special, but it has a tachyonic
excitation; on the other hand, the $\beta=0$ model is ghost and tachyon
free for $\kappa>0$.{]} Therefore, at the linearized level, the BHT
model is actually not a higher-derivative theory, so it escapes the
Ostragradski instability. The $\phi$ field part of the BHT action
reads \[
I_{BHT,\phi}=-\frac{1}{2\kappa}\int d^{3}x\,\left(\phi\Box\phi+\frac{1}{\kappa\beta}\phi^{2}\right),\]
 which again describes a single degree of freedom with the same mass
as $\sigma$. This is to be expected in this parity-invariant theory,
since $\sigma$ and $\phi$ are two helicity degrees of freedom of
the massive spin two field in three dimensions. Also, observe that
for $\phi$ to be a nonghost, $\kappa$ has to be negative.

For generic $\alpha$ and $\beta$, except for $2\alpha+\beta\neq0$,
(\ref{eq:phi_action}) describes a higher-derivative Pais-Uhlenbeck
\cite{pais} oscillator which can be rewritten in terms of simple
oscillators in the following way. Defining new fields as\[
\varphi_{1}\equiv\phi-\frac{\Box\phi}{m_{g}^{2}},\qquad\varphi_{2}\equiv\phi-\frac{\Box\phi}{m_{s}^{2}},\]
 (\ref{eq:phi_action}) becomes\begin{equation}
I_{\phi}=\frac{1}{64\kappa\left(2\alpha+\beta\right)^{2}}\int d^{3}x\,\left[\left(8\alpha+3\beta\right)^{2}\varphi_{1}\left(\Box-m_{s}^{2}\right)\varphi_{1}-\beta^{2}\varphi_{2}\left(\Box-m_{g}^{2}\right)\varphi_{2}\right],\label{eq:decoupled_phi_action}\end{equation}
 with $m_{g}$ given as above and $m_{s}$ as\[
m_{s}^{2}=\frac{1}{\kappa\left(8\alpha+3\beta\right)}.\]
 For $8\alpha+3\beta<0$, $\varphi_{1}$ is nontachyonic just like
$\varphi_{2}$, but unlike $\varphi_{2}$, it describes a ghostlike
excitation since its kinetic energy comes with the wrong sign.

\subsection{$2\alpha+\beta=0$ theory}

We have seen in the above discussion that the $2\alpha+\beta=0$ case
is a somewhat singular theory. If one naively takes the $\epsilon\equiv2\alpha+\beta\rightarrow0$
limit in (\ref{eq:decoupled_phi_action}), one gets \begin{align*}
I_{\phi} & =\frac{1}{8\kappa\epsilon}\int d^{3}x\,\left\{ \frac{\beta}{m_{g}^{2}}\left[\left(\Box-m_{g}^{2}\right)\phi\right]^{2}-4\epsilon\phi\left(\Box-m_{g}^{2}\right)\phi+O\left(\epsilon^{2}\right)\right\} ,\end{align*}
 which is a degenerate (equal mass) Pais-Uhlenbeck oscillator after
a divergent rescaling of $\phi$. But, more properly, suppose from
the onset at the level of the action, we set $2\alpha+\beta=0$ to
get (apart from the decoupled $\sigma$ field) \begin{align*}
I_{\phi} & =\frac{\beta}{2}\int d^{3}x\,\left(q\Box\phi-m_{g}^{2}q\phi\right).\end{align*}
 Variation with respect to $\phi$ gives a massive wave equation for
$q$, and \emph{vice versa}. But, these equations do not reveal the
ghost structure of the theory. So, let us define $q\equiv m_{g}^{2}\left(\Psi_{1}+\Psi_{2}\right)$,
$\phi\equiv\Psi_{1}-\Psi_{2}$, which turns the action to\begin{align*}
I & =\frac{m_{g}^{2}\beta}{2}\int d^{3}x\,\left[\left(\Psi_{1}\Box\Psi_{1}-m_{g}^{2}\Psi_{1}^{2}\right)-\left(\Psi_{2}\Box\Psi_{2}-m_{g}^{2}\Psi_{2}^{2}\right)\right].\end{align*}
 Since $\beta>0$, $\Psi_{2}$ is a ghost excitation. The Newtonian
limit of this theory is quite interesting: From the general tree-level
scattering amplitude computation given in \cite{gullu}, one sees
that as in the pure Einstein-Hilbert theory, the $2\alpha+\beta=0$
case has a vanishing Newtonian potential between static sources: the
spin-0 ghost excitation gives a repulsive component which cancels
the attractive one coming from the spin-2 part.

\subsection{Adding static sources}

Up to now, we have studied the free field spectrum of higher-derivative
gravity. Let us remedy this by adding matter with the usual gravity-matter
coupling:\[
I_{\text{source}}=\frac{1}{2}\int d^{3}x\, h_{\mu\nu}T^{\mu\nu}.\]
 In the case of a static source, $T^{00}=\rho\left(\vec{x}\right)$,
$T^{0i}=0$, $T^{ij}=0,$ (in a related context, we somewhat generalize
this in Appendix A), $I_{source}$ becomes \[
I_{\text{source}}=\frac{1}{2}\int d^{3}x\, N\rho\left(\vec{x}\right)=\frac{1}{2}\int d^{3}x\,\left(\frac{1}{\nabla^{2}}q+2\dot{N}_{L}-\frac{1}{\nabla^{2}}\ddot{\chi}\right)\rho\left(\vec{x}\right),\]
 where in the second equality, we have used the definition of $q$
in (\ref{eq:flat_gauge_inv_comb}). After dropping the boundary terms
and using the symmetry of the Green's function, we have\[
I_{\text{source}}=\frac{1}{2}\int d^{3}x\, q\frac{1}{\nabla^{2}}\rho.\]
 Redefining $\varphi\equiv\phi+\kappa\frac{1}{\nabla^{2}}\rho$ and
$\tilde{q}\equiv q+\kappa\rho$, the total action reduces to\begin{align*}
I & =\frac{1}{2}\int d^{3}x\,\left[\frac{1}{\kappa}\left(\varphi\tilde{q}-\kappa\varphi\rho+\sigma^{2}\right)+\left(2\alpha+\beta\right)\left(\tilde{q}-\Box\varphi\right)^{2}+\beta\left(\tilde{q}\Box\varphi-\kappa\rho\Box\varphi-\kappa\tilde{q}\rho+\kappa^{2}\rho^{2}+\sigma\Box\sigma\right)\right].\end{align*}
 Specifically, for $8\alpha+3\beta=0$, integrating out $\tilde{q}$,
one ends up with\[
I=\frac{1}{2}\int d^{3}x\,\left[\beta\left(\sigma\Box\sigma-m_{g}^{2}\sigma^{2}\right)-\frac{1}{\kappa}\left(\varphi\Box\varphi-m_{g}^{2}\varphi^{2}\right)+\varphi\rho\right].\]
 The last term is the interaction part which gives the attractive
(for $\kappa<0$) potential energy\begin{equation}
U=\frac{\kappa}{4}\int d^{2}x\,\rho_{1}\frac{1}{\nabla^{2}-m_{g}^{2}}\rho_{2}=\frac{\kappa}{8\pi}m_{1}m_{2}\text{\ensuremath{K_{0}}}\left(m_{g}r\right),\label{eq:Potential_source}\end{equation}
 where we took point sources, $\rho_{1}\left(\vec{x}\right)=m_{1}\delta^{\left(2\right)}\left(\vec{x}-\vec{x}_{1}\right)$,
$\rho_{2}\left(\vec{x}\right)=m_{2}\delta^{\left(2\right)}\left(\vec{x}-\vec{x}_{2}\right)$,
and $\text{\ensuremath{K_{0}}}$ is the modified Bessel function.
This result matches that of \cite{gullu}.

\subsection{Weak field approximation}

It is also highly instructive to capture some of the above results
from the nonlinear theory (\ref{eq:Non-linear_action}). But, even
in the circularly symmetric case, nontrivial exact solutions for which
$g_{00}\ne g^{rr}$ are not known, and we have not been able to find
one. Nevertheless, since we just need the weak field approximation,
we can do the following: The ansatz \textbf{\[
ds^{2}=-f\left(r\right)dt^{2}+\frac{b^{2}\left(r\right)}{f\left(r\right)}dr^{2}+r^{2}d\theta^{2},\]
 }can be inserted into the action (\ref{eq:Non-linear_action}), which
is to be varied with respect $f\left(r\right)$ and $b\left(r\right)$
{[}See the details of this Weyl trick in \cite{dt1}{]} . For the
sake of simplicity, let us just consider the BHT theory. Then, an
approximate solution can be found by setting $f\left(r\right)=1+\int^{r}dr\, a\left(r\right)$,
$b\left(r\right)=1+\int^{r}dr\, v\left(r\right)$, where $a$ and
$v$ are small. At the first order, we have\begin{equation}
\frac{4}{\kappa}v+2\beta v^{\prime\prime}+2\beta a^{\prime\prime}+r\beta a^{\prime\prime\prime}=0,\label{eq:w_A_eqn1}\end{equation}
 \begin{equation}
\beta r^{2}a^{\prime\prime}+\frac{2}{\kappa}r^{2}a+2r\beta v'-2\beta v=0.\label{eq:dif_of_w_A_eqns}\end{equation}
 Here, $^{\prime}$ denotes differentiation with respect to $r$.
$v$ can be determined as $v=a+\frac{r}{2}a^{\prime}.$ Putting it
back to (\ref{eq:dif_of_w_A_eqns}) gives\begin{align}
r^{2}a^{\prime\prime}+ra^{\prime}-a\left(m_{g}^{2}r^{2}+1\right) & =0,\label{eq:A_eqn2}\end{align}
 which is solved by $a\left(r\right)=c_{1}\text{\ensuremath{I_{1}}}\left(m_{g}r\right)+c_{2}\text{\ensuremath{K_{1}}}\left(m_{g}r\right)$.
Recall that $g_{00}\approx-1-\int^{r}dr\, a\left(r\right),$ and $g_{rr}\approx1+\int^{r}dr\,\left(2v\left(r\right)-a\left(r\right)\right)$.
Thus, for decaying fields $c_{1}$ vanishes, and the metric components
become \[
g_{00}\approx-1+c\text{\ensuremath{K_{0}}}\left(m_{g}r\right),\qquad g_{rr}\approx1+d\text{\ensuremath{K_{1}}}\left(m_{g}r\right),\]
 where $c$ and $d$ are constants related to the mass of the source.
This is consistent with our earlier result (\ref{eq:Potential_source}).

\subsection{Higher-derivative gravity plus a Chern-Simons term}

We will now extend the preceding discussion in flat space by adding
a gravitational Chern-Simons term \cite{djt}\begin{equation}
I=\int d^{3}x\,\sqrt{-g}\left[\frac{1}{\kappa}R+\alpha R^{2}+\beta R_{\mu\nu}^{2}-\frac{1}{2\mu}\epsilon^{\lambda\mu\nu}\Gamma_{\phantom{\rho}\lambda\sigma}^{\rho}\left(\partial_{\mu}\Gamma_{\phantom{\sigma}\rho\nu}^{\sigma}+\frac{2}{3}\Gamma_{\phantom{\sigma}\mu\beta}^{\sigma}\Gamma_{\phantom{\sigma}\nu\rho}^{\beta}\right)\right],\label{eq:CS_non-linear_action}\end{equation}
 where $\epsilon_{012}=1$, and $\mu$ is the Chern-Simons coupling
with an arbitrary sign. {[}Without the $\alpha$, $\beta$ terms,
but with a Pauli-Fierz mass term, canonical analysis was carried out
in \cite{dt2,st}{]} Linearization of the Chern-Simons part yields\[
I_{CS}=-\frac{1}{2\mu}\int d^{3}x\,\epsilon_{\mu\alpha\beta}\mathcal{G}_{L}^{\alpha\nu}\partial^{\mu}h_{\phantom{\beta}\nu}^{\beta}=\frac{1}{2\mu}\int d^{3}x\,\sigma\left(q+\Box\phi\right).\]
 The total action in terms of the gauge-invariant combinations becomes\begin{align*}
I & =\frac{1}{2}\int d^{3}x\,\left[\frac{1}{\kappa}\left(\phi q+\sigma^{2}\right)+\left(2\alpha+\beta\right)\left(q-\Box\phi\right)^{2}+\beta\left(q\Box\phi+\sigma\Box\sigma\right)+\frac{1}{\mu}\sigma\left(q+\Box\phi\right)\right].\end{align*}
 Assuming that $2\alpha+\beta\ne0$, $q$ can be eliminated to yield
the action\begin{align*}
I & =\frac{1}{2}\int d^{3}x\,\left\{ \beta\left[\sigma\Box\sigma+\left(\frac{1}{\kappa\beta}-\frac{1}{4\mu^{2}\beta\left(2\alpha+\beta\right)}\right)\sigma^{2}\right]+\left[\frac{1}{\mu}+\frac{\left(4\alpha+\beta\right)}{2\mu\left(2\alpha+\beta\right)}\right]\sigma\Box\phi-\frac{1}{2\kappa\mu\left(2\alpha+\beta\right)}\sigma\phi\right.\\
 & \phantom{=\frac{1}{2}\int d^{3}x\,}\left.+\frac{1}{\kappa}\left[\frac{\beta\kappa\left(8\alpha+3\beta\right)}{4\left(2\alpha+\beta\right)}\left(\Box\phi\right)^{2}+\frac{\left(4\alpha+\beta\right)}{2\left(2\alpha+\beta\right)}\phi\Box\phi-\frac{1}{4\kappa\left(2\alpha+\beta\right)}\phi^{2}\right]\right\} .\end{align*}
 For generic $\alpha$,$\beta$ one can diagonalize this action, but
it is rather cumbersome and not particularly illuminating, so we just
consider the $8\alpha+3\beta=0$ case, \begin{align*}
I_{BHT-CS} & =\frac{\beta}{2}\int d^{3}x\,\left\{ \left[\sigma\Box\sigma-\left(m_{g}^{2}+\frac{1}{\mu^{2}\beta^{2}}\right)\sigma^{2}\right]+\frac{2m_{g}^{2}}{\beta\mu}\sigma\phi+m_{g}^{2}\left(\phi\Box\phi-m_{g}^{2}\phi^{2}\right)\right\} .\end{align*}
 To decouple the $\sigma$, $\phi$ fields, one possible route is
to take the Fourier transform of the fields, put the Lagrangian in
a matrix form, and then diagonalize the matrix. This procedure yields\[
I_{BHT-CS}=\frac{\beta}{2}\int d^{3}x\,\left(\Psi_{+}\Box\Psi_{+}-m_{+}^{2}\Psi_{+}^{2}+\Psi_{-}\Box\Psi_{-}-m_{-}^{2}\Psi_{-}^{2}\right),\]
 where the masses read\[
m_{\pm}^{2}=m_{g}^{2}+\frac{1}{2\mu^{2}\beta^{2}}\pm\frac{1}{\mu\beta}\sqrt{m_{g}^{2}+\frac{1}{4\mu^{2}\beta^{2}}},\]
 and the new fields are defined as\[
\left(\begin{array}{c}
\Psi_{-}\\
\Psi_{+}\end{array}\right)=\left[\begin{array}{cc}
N_{+}\qquad & \left(m_{+}^{2}-m_{g}^{2}\right)N_{+}\\
N_{-}\qquad & \left(m_{-}^{2}-m_{g}^{2}\right)N_{-}\end{array}\right]\left(\begin{array}{c}
\sigma\\
m_{g}\phi\end{array}\right),\qquad N_{\pm}=\sqrt{1+\left[\frac{\mu\beta}{m_{g}}\left(m_{\pm}^{2}-m_{g}^{2}\right)\right]^{2}}.\]
 $m_{\pm}$ agree with those of \cite{bht,andringa}. As the $+2$
and $-2$ helicity modes have different masses, it is a parity-violating
theory as expected. In the $\beta\rightarrow0$ limit, which is the
topologically massive gravity with a single degree of freedom \cite{djt},
$m_{+}$ diverges and drops out, $m_{-}=-\left|\mu\right|/\kappa$.

\section{Higher-derivative spin-2 in a de Sitter background\label{sec:dS_space}}

Now, we will study the canonical structure of the higher-derivative
theory in an (anti)-de Sitter background defined by the action\[
I=\int d^{3}x\,\sqrt{-g}\left[\frac{1}{\kappa}\left(R-2\Lambda_{0}\right)+\alpha R^{2}+\beta R_{\mu\nu}^{2}\right],\]
whose linearization about an (a)dS background yields\[
I=-\frac{1}{2}\int d^{3}x\,\sqrt{-\bar{g}}\, h_{\mu\nu}\left[a\mathcal{G}_{L}^{\mu\nu}+\left(2\alpha+\beta\right)\left(\bar{g}^{\mu\nu}\Box-\nabla^{\mu}\nabla^{\nu}+\frac{2}{\ell^{2}}\bar{g}^{\mu\nu}\right)R_{L}+\beta\left(\Box\mathcal{G}_{L}^{\mu\nu}-\frac{1}{\ell^{2}}\bar{g}^{\mu\nu}R_{L}\right)\right],\]
 where $a\equiv\frac{1}{\kappa}+\frac{12}{\ell^{2}}\alpha+\frac{2}{\ell^{2}}\beta$,
and $1/\ell^{2}$ is the cosmological constant which is related to
$\alpha$, $\beta$, $\kappa$ and the bare cosmological constant
$\Lambda_{0}$ of the full theory as $\frac{1}{\ell^{2}}=\frac{1}{4\kappa\left(3\alpha+\beta\right)}\left[1\pm\sqrt{1-8\kappa\Lambda_{0}\left(3\alpha+\beta\right)}\right]$
\cite{dt3}. For the sake of simplicity, we will consider the background
to be a de Sitter spacetime, but since our results will be analytic
in $\ell$, in the final expressions one can take $\ell\rightarrow i\ell$
to obtain the results in anti-de Sitter spacetime. {[}To keep the
signature intact, one also needs to Wick rotate a space coordinate{]}.
For dS, we take the metric, $\bar{g}_{\mu\nu}$, with which all the
covariant derivatives and raising-lowering operations should be made,
to be in the \[
ds^{2}=\frac{\ell^{2}}{t^{2}}\left(-dt^{2}+dx^{2}+dy^{2}\right),\]
 and define the perturbation as\[
g_{\mu\nu}=\frac{\ell^{2}}{t^{2}}\eta_{\mu\nu}+h_{\mu\nu}.\]
 Linearized forms of Einstein and Ricci tensors, and Ricci scalar
are given as\begin{gather}
\mathcal{G}_{\mu\nu}^{L}=R_{\mu\nu}^{L}-\frac{1}{2}\bar{g}_{\mu\nu}R_{L}-\frac{2}{\ell^{2}}h_{\mu\nu},\nonumber \\
R_{\mu\nu}^{L}=\frac{1}{2}\left(\nabla^{\sigma}\nabla_{\mu}h_{\nu\sigma}+\nabla^{\sigma}\nabla_{\nu}h_{\mu\sigma}-\Box h_{\mu\nu}-\nabla_{\mu}\nabla_{\nu}h\right),\qquad R_{L}=\nabla_{\alpha}\nabla_{\beta}h^{\alpha\beta}-\Box h-\frac{2}{\ell^{2}}h,\label{eq:ds_lin_tensors}\end{gather}
 where $\Box\equiv\nabla_{\mu}\nabla^{\mu}=\frac{t^{2}}{\ell^{2}}\eta^{\mu\nu}\nabla_{\mu}\nabla_{\nu}$.
Decomposition of $h_{\mu\nu}$ into {}``spatial'' tensor $h_{ij}$,
{}``spatial'' vector $h_{0i}$, and {}``scalar'' $h_{00}$ is\begin{align*}
h_{ij} & \equiv\frac{\ell^{2}}{t^{2}}\left[\left(\delta_{ij}+\hat{\nabla}_{i}\hat{\nabla}_{j}\right)\phi-\hat{\nabla}_{i}\hat{\nabla}_{j}\chi+\left(\tilde{\epsilon}_{i}^{\phantom{i}k}\hat{\nabla}_{k}\hat{\nabla}_{j}+\tilde{\epsilon}_{j}^{\phantom{j}k}\hat{\nabla}_{k}\hat{\nabla}_{i}\right)\xi\right]\\
 & =\frac{\ell^{2}}{t^{2}}\left[\left(\delta_{ij}+\hat{\nabla}_{i}\hat{\nabla}_{j}\right)\phi-\hat{\nabla}_{i}\hat{\nabla}_{j}\chi+\frac{t^{2}}{\ell^{2}}\left(\tilde{\epsilon}_{ik}\hat{\nabla}_{k}\hat{\nabla}_{j}+\tilde{\epsilon}_{jk}\hat{\nabla}_{k}\hat{\nabla}_{i}\right)\xi\right],\\
h_{0i} & \equiv\frac{\ell^{2}}{t^{2}}\left(-\tilde{\epsilon}_{i}^{\phantom{i}k}\nabla_{k}\eta+\partial_{i}N_{L}\right)=\frac{\ell^{2}}{t^{2}}\left(-\frac{t^{2}}{\ell^{2}}\tilde{\epsilon}_{ij}\nabla_{j}\eta+\partial_{i}N_{L}\right),\\
h_{00} & \equiv\frac{\ell^{2}}{t^{2}}N,\end{align*}
where $\hat{\nabla}_{i}\equiv\nabla_{i}/\sqrt{-\nabla_{k}^{2}}$ and
the covariant derivative is for two-dimensional space with metric
$\gamma_{ij}=\frac{\ell^{2}}{t^{2}}\delta_{ij}$. Since the two-dimensional
space is flat, then $\nabla_{i}\rightarrow\partial_{i}$ and $\hat{\partial}_{i}\equiv\partial_{i}/\sqrt{-\partial_{k}^{2}}$.
$\tilde{\epsilon}_{ik}$ is the Levi-Civita tensor for two-dimensional
space, which is related with the corresponding tensor density $\epsilon_{ik}$
by\[
\tilde{\epsilon}_{ik}=\sqrt{\gamma}\epsilon_{ik}\quad\Rightarrow\quad\tilde{\epsilon}_{ik}=\frac{\ell^{2}}{t^{2}}\epsilon_{ik}.\]
The convention for $\epsilon_{ik}$ is $\epsilon_{12}=1$ (the convention
for Levi-Civita tensor density for the upper indices is $\epsilon^{12}=1$
naturally with the induced metric). As a result, the final form of
the decomposition is\begin{align*}
h_{ij} & =\frac{\ell^{2}}{t^{2}}\left[\left(\delta_{ij}+\hat{\partial}_{i}\hat{\partial}_{j}\right)\phi-\hat{\partial}_{i}\hat{\partial}_{j}\chi+\left(\epsilon_{ik}\hat{\partial}_{k}\hat{\partial}_{j}+\epsilon_{jk}\hat{\partial}_{k}\hat{\partial}_{i}\right)\xi\right],\\
h_{0i} & =\frac{\ell^{2}}{t^{2}}\left(-\epsilon_{ij}\partial_{j}\eta+\partial_{i}N_{L}\right),\qquad h_{00}=\frac{\ell^{2}}{t^{2}}N,\end{align*}
with the convention for Levi-Civita tensor \emph{density} $\epsilon_{12}=1$.
Here, all the spatial indices are raised and lowered by $\delta_{ij}$.
A further note on this specific choice of decomposition is about the
$\ell^{2}/t^{2}$ coefficients: With this coefficients, at every step
the flat space limit $\ell\rightarrow\infty$, $\ell/t\rightarrow1$
will be clear. 

Unlike the flat space case, $\phi$ is not gauge invariant anymore.
In fact, under the gauge transformations $\delta_{\zeta}h_{\mu\nu}=\nabla_{\mu}\zeta_{\nu}+\nabla_{\nu}\zeta_{\mu}$
where $\zeta_{\mu}$ can be decomposed as $\zeta_{\mu}=\left(\zeta_{0},-\epsilon_{ij}\partial_{j}\zeta+\partial_{i}\kappa\right)$,
the components of $h_{\mu\nu}$ transform as\begin{align*}
\delta_{\zeta}\phi & =2\frac{t}{\ell^{2}}\zeta_{0},\qquad\delta_{\zeta}\chi=2\frac{t^{2}}{\ell^{2}}\left(\partial_{i}^{2}\kappa+\frac{1}{t}\zeta_{0}\right),\qquad\delta_{\zeta}\xi=\frac{t^{2}}{\ell^{2}}\partial_{i}^{2}\zeta,\\
\delta_{\zeta}\eta & =\frac{t^{2}}{\ell^{2}}\left(\dot{\zeta}+\frac{2}{t}\zeta\right),\qquad\delta_{\zeta}N_{L}=\frac{t^{2}}{\ell^{2}}\left(\dot{\kappa}+\zeta_{0}+\frac{2}{t}\kappa\right),\qquad\delta_{\zeta}N=2\frac{t^{2}}{\ell^{2}}\left(\dot{\zeta}_{0}+\frac{1}{t}\zeta_{0}\right).\end{align*}
 Again, from the linearized Bianchi identity, $\nabla_{\mu}\mathcal{G}_{L}^{\mu\nu}=0$,
we know that there should be three independent gauge-invariant combinations
constructed out of the (derivatives of) six scalar fields. By inspection,
one can find these combinations, but the quickest way would be to
look at the independent components of the gauge-invariant tensor $\mathcal{G}_{L}^{\mu\nu}$.
This led us to the following \emph{four} gauge-invariant functions:
\begin{align*}
f & \equiv\frac{\ell}{t}\left[\phi-\frac{2}{t}N_{L}+\frac{1}{t}\frac{1}{\nabla^{2}}\left(\dot{\phi}+\dot{\chi}-\frac{2}{t}N\right)\right],\qquad p\equiv\frac{\ell}{t}\left(\dot{\phi}-\frac{1}{t}N\right),\\
q & \equiv\frac{\ell}{t}\left[\nabla^{2}N+\ddot{\chi}-2\nabla^{2}\dot{N}_{L}-\frac{1}{t}\left(\dot{N}-2\nabla^{2}N_{L}+\dot{\chi}\right)+\frac{2}{t^{2}}N\right],\qquad\sigma\equiv\frac{\ell}{t}\left(\dot{\xi}-\nabla^{2}\eta\right),\end{align*}
 and a relation between them coming from the Bianchi identity\begin{equation}
t\nabla^{2}\left(\dot{f}-p+\frac{f}{t}\right)-\dot{p}-q=0.\label{eq:Bianchi_id}\end{equation}
 In terms of these, the components of the linearized Einstein tensor
can be found as\begin{gather*}
\mathcal{G}_{00}^{L}=-\frac{t}{2\ell}\nabla^{2}f,\qquad\mathcal{G}_{0i}^{L}=-\frac{t}{2\ell}\left(\partial_{i}p+\epsilon_{ik}\partial_{k}\sigma\right),\\
\mathcal{G}_{ij}^{L}=-\frac{t}{2\ell}\left[\left(\delta_{ij}+\hat{\partial}_{i}\hat{\partial}_{j}\right)q-\hat{\partial}_{i}\hat{\partial}_{j}\dot{p}-\left(\epsilon_{ik}\hat{\partial}_{k}\hat{\partial}_{j}+\epsilon_{jk}\hat{\partial}_{k}\hat{\partial_{i}}\right)\dot{\sigma}\right].\end{gather*}
 The linearized curvature scalar follows as\[
R_{L}=\frac{t^{3}}{\ell^{3}}\left(q-\nabla^{2}f+\dot{p}\right)=\frac{t^{4}}{\ell^{3}}\nabla^{2}\left(\dot{f}-p\right),\]
 where in the second line we used the Bianchi identity.

Using the above, the Einstein-Hilbert action can be reduced to the
following form:\[
I_{EH}=-\frac{a}{2}\int d^{3}x\,\sqrt{-\bar{g}}\, h_{\mu\nu}\mathcal{G}_{L}^{\mu\nu}=\frac{a}{2}\int d^{3}x\,\left[\frac{\ell^{2}}{t^{2}}fR_{L}+\frac{t}{\ell}\left(f\nabla^{2}f+p^{2}+\sigma^{2}\right)\right].\]
 As in the flat space case, computations get a lot simpler if the
higher-derivative parts of the Lagrangian are organized in such a
way that $h_{\mu\nu}$ is replaced by some gauge-invariant combinations.
This can be done again upon use of the self-adjointness of the involved
operators as follows:\[
I_{2\alpha+\beta}=-\frac{\left(2\alpha+\beta\right)}{2}\int d^{3}x\,\sqrt{-\bar{g}}h_{\mu\nu}\left(\bar{g}^{\mu\nu}\Box-\nabla^{\mu}\nabla^{\nu}+\frac{2}{\ell^{2}}\bar{g}^{\mu\nu}\right)R_{L}=\frac{\left(2\alpha+\beta\right)}{2}\int d^{3}x\,\sqrt{-\bar{g}}R_{L}^{2}.\]
 For the $\beta$ term, one has \begin{align*}
I_{\beta} & =-\frac{\beta}{2}\int d^{3}x\,\sqrt{-\bar{g}}h_{\mu\nu}\left(\Box\mathcal{G}_{L}^{\mu\nu}-\frac{1}{\ell^{2}}\bar{g}^{\mu\nu}R_{L}\right)=-\frac{\beta}{2}\int d^{3}x\,\sqrt{-\bar{g}}\left[\left(\Box h_{\mu\nu}\right)\mathcal{G}_{L}^{\mu\nu}-\frac{1}{\ell^{2}}hR_{L}\right].\end{align*}
 After organizing $R_{\mu\nu}^{L}$ (\ref{eq:ds_lin_tensors}) into
a form where the indices $\mu$ and $\nu$ in the covariant derivatives
stay at the far left, and using the Bianchi identity, $\nabla_{\mu}\mathcal{G}_{L}^{\mu\nu}=0$,
one arrives at \[
I_{\beta}=-\frac{\beta}{2}\int d^{3}x\,\sqrt{-\bar{g}}\left(-2\mathcal{G}_{\mu\nu}^{L}\mathcal{G}_{L}^{\mu\nu}+\frac{1}{2}R_{L}^{2}+\frac{2}{\ell^{2}}h_{\mu\nu}\mathcal{G}_{L}^{\mu\nu}\right).\]
 Note that, had we not done this and instead computed $h_{\mu\nu}\Box\mathcal{G}_{L}^{\mu\nu}$
directly, putting the result into an explicitly gauge-invariant form
would be somewhat time-consuming. Not worrying about the correct canonical
dimensions for the fields, one can collect all the parts computed
above to end up with\begin{align*}
I & =\frac{1}{2}\int d^{3}x\,\left\{ \left(a+\frac{2\beta}{\ell^{2}}\right)\left[\frac{\ell^{2}}{t^{2}}fR_{L}+\frac{t}{\ell}\left(f\nabla^{2}f+p^{2}+\sigma^{2}\right)\right]+\left(2\alpha+\beta\right)\frac{\ell^{3}}{t^{3}}R_{L}^{2}\right.\\
 & \phantom{=\frac{1}{2}\int d^{3}x\,}\left.+\beta\frac{t^{3}}{\ell^{3}}\left[\dot{\sigma}^{2}+\sigma\nabla^{2}\sigma+\dot{p}^{2}+p\nabla^{2}p+\left(\nabla^{2}f\right)^{2}\right.\right.\\
 & \phantom{=\frac{1}{2}\int d^{3}x\,+\beta\frac{t^{3}}{\ell^{3}}}\left.\left.+\frac{\ell^{3}}{t^{3}}R_{L}\nabla^{2}f-\frac{\ell^{3}}{t^{3}}R_{L}\dot{p}-\dot{p}\nabla^{2}f\right]\right\} .\end{align*}
 The flat space limit of this action gives (\ref{eq:flat_total_action}).
In this form, not all the fields are independent: After defining $\varphi\equiv\nabla^{2}f$,
and using the Bianchi identity (\ref{eq:Bianchi_id}), we can further
simplify the action to\begin{align}
I & =\frac{1}{2}\int d^{3}x\,\left\{ \left(a+\frac{2\beta}{\ell^{2}}\right)\frac{t}{\ell}\left(-tp\varphi+p^{2}\right)+\left(2\alpha+\beta\right)\frac{t^{5}}{\ell^{3}}\left(\dot{\varphi}-\nabla^{2}p\right)^{2}\right.\nonumber \\
 & \phantom{=\frac{1}{2}\int d^{3}x\,}\left.+\beta\frac{t^{3}}{\ell^{3}}\left(\dot{p}^{2}-p\nabla^{2}p-\varphi^{2}-t\varphi\nabla^{2}p-t\dot{p}\dot{\varphi}-\varphi\dot{p}\right)\right\} +I_{\sigma},\label{eq:total_action_in_3_variable}\end{align}
 where the $\sigma$ field decouples from the rest \begin{equation}
I_{\sigma}=\frac{1}{2}\int d^{3}x\,\left[\beta\frac{t^{3}}{\ell^{3}}\left(\dot{\sigma}^{2}+\sigma\nabla^{2}\sigma\right)+\left(a+\frac{2\beta}{\ell^{2}}\right)\frac{t}{\ell}\sigma^{2}\right].\label{eq:sigma_action}\end{equation}
 For vanishing $\alpha$ and $\beta$, cosmological Einstein theory
does not have any propagating degrees of freedom just like its flat
space partner. For generic $\alpha$ and $\beta$, there are 3 degrees
of freedom. Recall that a minimally coupled scalar field with the
correct canonical dimension is in the following form:\[
I=-\frac{1}{2}\int d^{3}x\,\sqrt{-g}\left(\partial_{\mu}\Phi\partial^{\mu}\Phi+m^{2}\Phi^{2}\right)=-\frac{1}{2}\int d^{3}x\,\left\{ \frac{\ell}{t}\left[-\dot{\Phi}^{2}+\left(\partial_{i}\Phi\right)^{2}\right]+\frac{\ell^{3}}{t^{3}}m^{2}\Phi^{2}\right\} .\]
 Therefore, after rescaling $\sigma\rightarrow\frac{\ell^{2}}{t^{2}}\sigma$
in (\ref{eq:sigma_action}), one finds the mass of the $\sigma$ field
as\begin{equation}
m_{g}^{2}=-\frac{a}{\beta}-\frac{2}{\ell^{2}}=-\frac{1}{\kappa\beta}-\frac{12\alpha}{\ell^{2}\beta}-\frac{4}{\ell^{2}}.\label{eq:dS_m_g}\end{equation}
 For generic $\alpha$ and $\beta$, unlike the flat space case, diagonalizing
the $\varphi$, $p$ action is highly nontrivial. But, there are various
ways to see the basic oscillators in this model. One such method is
to Fourier transform the fields just in the $\vec{x}$ space and then
consider the zero two-momentum limit. That would be equivalent to
dropping the $\nabla^{2}$ terms in the action. Note that this construction
does not change the number of degrees of freedom, of course as long
as $\nabla^{2}\left(\text{field}\right)$ is not the lowest order
term. Another way is to directly study the equations of motion. We
shall employ both of these methods below.

\subsection{Masses from the nonrelativistic limit}

Apart from the decoupled $\sigma$ part, the generic $\alpha$, $\beta$
theory (\ref{eq:total_action_in_3_variable}) reads in the nonrelativistic
limit as\[
I=\frac{1}{2}\int d^{3}x\,\left[\left(a+\frac{2\beta}{\ell^{2}}\right)\frac{t}{\ell}\left(-tp\varphi+p^{2}\right)+\left(2\alpha+\beta\right)\frac{t^{5}}{\ell^{3}}\dot{\varphi}^{2}+\beta\frac{t^{3}}{\ell^{3}}\left(\dot{p}^{2}-\varphi^{2}-t\dot{p}\dot{\varphi}-\varphi\dot{p}\right)\right].\]
 To decouple the fields, first note that $2\alpha+\beta=\frac{\beta}{4}+\frac{8\alpha+3\beta}{4}$,
and rescale $\varphi$ as $\varphi\rightarrow\frac{1}{t}\varphi$
to get the action\begin{align*}
I & =\frac{1}{2}\int d^{3}x\,\left[\left(a+\frac{2\beta}{\ell^{2}}\right)\frac{t}{\ell}\left(-p\varphi+p^{2}\right)+\frac{\beta}{4}\frac{t^{3}}{\ell^{3}}\left(\dot{\varphi}^{2}-\frac{\varphi^{2}}{t^{2}}+4\dot{p}^{2}-4\dot{p}\dot{\varphi}\right)\right.\\
 & \phantom{=\frac{1}{2}\int d^{3}x\,}+\left.\frac{\left(8\alpha+3\beta\right)}{4}\frac{t^{3}}{\ell^{3}}\left(\dot{\varphi}^{2}+\frac{3\varphi^{2}}{t^{2}}\right)\right].\end{align*}
 Then, define a new field as $\Phi\equiv\varphi-2p$, which leads
to the decoupled actions for the $\Phi$ and $\varphi$ fields. As
the spin-2 helicity partner of the $\sigma$ field, the $\Phi$ action
is exactly like the $\sigma$ action with the same mass $m_{g}$ (\ref{eq:dS_m_g});
\[
I_{\Phi}=\frac{\beta}{8}\int d^{3}x\,\left[\frac{t^{3}}{\ell^{3}}\dot{\Phi}^{2}+\frac{t}{\ell}\left(\frac{a}{\beta}+\frac{2}{\ell^{2}}\right)\Phi^{2}\right],\]
 and the spin-0 mode has the action \[
I_{\varphi}=\frac{\left(8\alpha+3\beta\right)}{8}\int d^{3}x\,\left[\frac{t^{3}}{\ell^{3}}\dot{\varphi}^{2}-\frac{1}{\left(8\alpha+3\beta\right)}\frac{t}{\ell}\left(a-\frac{24\alpha}{\ell^{2}}-\frac{6\beta}{\ell^{2}}\right)\varphi^{2}\right],\]
 which after putting into the canonical form by rescaling $\varphi\rightarrow\frac{\ell^{2}}{t^{2}}\varphi$
yields the mass \[
m_{s}^{2}=\frac{1}{\kappa\left(8\alpha+3\beta\right)}-\frac{4}{\ell^{2}}\left(\frac{3\alpha+\beta}{8\alpha+3\beta}\right).\]
 In the $8\alpha+3\beta=0$ case, the $\varphi$ field freezes out
and $m_{g}^{2}$ matches the result of \cite{bht} obtained with the
help of an auxiliary field, not via canonical analysis. For generic
$\alpha$ and $\beta$, in accordance with the analysis of \cite{bht},
one can introduce two auxiliary fields to rewrite the action (\ref{eq:Non-linear_action}),
but decoupling of the scalar mode from the spin-2 mode is not immediately
clear. This is done in Appendix B.

\subsection{Equations of motions in the BHT case}

The above nonrelativistic analysis reveals the canonical structure
of the generic $\alpha$, $\beta$ theory. But here let us consider
the relativistic equations of motion for the $8\alpha+3\beta=0$ case.
Dropping the $\sigma$ field in (\ref{eq:total_action_in_3_variable}),
we have\begin{align*}
I & =\frac{\beta}{2}\int d^{3}x\,\left\{ m_{g}^{2}\frac{t}{\ell}\left(tp\varphi-p^{2}\right)+\frac{t^{5}}{4\ell^{3}}\left(\dot{\varphi}-\nabla^{2}p\right)^{2}\right.\\
 & \phantom{=\frac{1}{2}\int d^{3}x\,}\left.+\frac{t^{3}}{\ell^{3}}\left(\dot{p}^{2}-p\nabla^{2}p-\varphi^{2}-t\varphi\nabla^{2}p-t\dot{p}\dot{\varphi}-\varphi\dot{p}\right)\right\} .\end{align*}
 It appears that there are 2 degrees of freedom in this action (which
would conflict our earlier result, and the result of \cite{bht}),
but this is a red herring, there is only a single degree of freedom.
A quick way to see this is to look at the Hessian matrix, $\mathcal{H}=\frac{\partial^{2}\mathcal{L}}{\partial\dot{q}_{i}\partial\dot{q}_{j}}$,
\[
\mathcal{H}=\frac{\beta t^{3}}{4\ell^{3}}\left(\begin{array}{cc}
t^{2} & -2t\\
-2t & 4\end{array}\right).\]
 Since $\det\mathcal{H}=0$, there is a constraint in the model. Therefore,
{}``velocities'' $\dot{\varphi}$ and $\dot{p}$ cannot be separately
expressed in terms of the canonical momenta\[
\Pi_{\varphi}\equiv\frac{\partial\mathcal{L}}{\partial\dot{\varphi}}=\frac{\beta t^{5}}{4\ell^{3}}\left(\dot{\varphi}-\nabla^{2}p-\frac{2}{t}\dot{p}\right),\qquad\Pi_{p}\equiv\frac{\partial\mathcal{L}}{\partial\dot{p}}=\frac{\beta t^{3}}{2\ell^{3}}\left(2\dot{p}-t\dot{\varphi}-\varphi\right).\]
 One can use the Dirac's constraint analysis method to obtain the
Hamiltonian for this singular Lagrangian, but here it suffices to
consider just the field equations. Taking the variations with respect
to $\varphi$ and $p$ yield\[
\delta\varphi:\quad\frac{m_{g}^{2}t^{2}}{\ell}p-\frac{t^{3}}{\ell^{3}}\left(2\varphi+t\nabla^{2}p+\dot{p}\right)-\frac{1}{2\ell^{3}}\partial_{0}\left[t^{5}\left(\dot{\varphi}-\nabla^{2}p\right)-2t^{4}\dot{p}\right]=0,\]
 and\[
\delta p:\quad\frac{m_{g}^{2}t}{\ell}\left(t\varphi-2p\right)-\frac{t^{5}}{2\ell^{3}}\nabla^{2}\left(\dot{\varphi}-\nabla^{2}p+\frac{4}{t^{2}}p+\frac{2}{t}\varphi\right)-\frac{1}{\ell^{3}}\partial_{0}\left[t^{3}\left(2\dot{p}-t\dot{\varphi}-\varphi\right)\right]=0.\]
 By inspection, and with a hint from the field equations which give
$R_{L}=0$, one observes that $\dot{\varphi}=\nabla^{2}p$ and the
other equation reduces to\[
\frac{\ell}{t}\left(-\ddot{\varphi}-\frac{1}{t}\dot{\varphi}+\nabla^{2}\varphi\right)-\frac{\ell^{3}}{t^{3}}\left(m_{g}^{2}-\frac{1}{\ell^{2}}\right)\varphi=0,\]
 which is not yet in the canonical wave equation form in dS. To put
in the canonical form, $\left(\Box-m^{2}\right)\phi=0$, rescale $\varphi\rightarrow\varphi/t$
to obtain\[
\frac{\ell}{t}\left(-\ddot{\varphi}+\frac{1}{t}\dot{\varphi}+\nabla^{2}\varphi\right)-\frac{\ell^{3}}{t^{3}}m_{g}^{2}\varphi=0,\Rightarrow\left(\Box-m_{g}^{2}\right)\varphi=0,\]
 which is exactly like the $\sigma$ field.

\section{Conclusions}

We have studied the canonical structure of the linearized quadratic
gravity models in an explicitly gauge-invariant way for both flat
and dS backgrounds in three dimensions. In flat spacetime, the general
action is decoupled into three harmonic oscillators. After considering
the signs and various limits of the parameters $\kappa$, $\alpha$,
$\beta$, the BHT case is singled out as the unique unitary and nontachyonic
theory (namely, a regular massive free spin-2 field, not a higher-time
derivative one), while the others are all higher-derivative Pais-Uhlenbeck
oscillators. Sources are also added to the theory, and Newtonian potentials
for both static and spinning particles are calculated. Moreover, we
have computed the weak field limit of the circularly symmetric spacetime.
We extended our flat space analysis to include the gravitational Chern-Simons
term and investigated the oscillator structure for the BHT limit:
We have seen that in this limit the oscillators decouple with different
masses, violating parity as expected. In dS, we have also found the
most general action in terms of three gauge-invariant functions constructed
from the (derivatives of the) components of the metric perturbation
and carried out the decoupling of the fields in the nonrelativistic
limit at the level of the action and in a relativistic form at the
level of the field equations. For future work, to go beyond the free
field level and introduce nonlinearities, such as $O\left(h^{3}\right)$
and interactions, our gauge-invariant actions will be of great use.
Another interesting point about the models that we discussed here
is that, especially in (anti)-de Sitter backgrounds, for certain tuned
values of the parameters novel phenomena such as partial masslessness
or chiral gravity arise. These topics will be addressed in a separate
work.

\section*{\label{ackno} Acknowledgments}

I.G. and B.T. are partially supported by the T{Ü}B\.{I}TAK Kariyer
Grant No. 104T177. T.Ç.\c{S}. is supported by a T{Ü}B\.{I}TAK Ph.D.
Scholarship.

\section*{Appendix A: Spinning masses}

It is also of some interest to understand how spinning point particles
interact in the generic higher-derivative model. This can be done
as follows: First, note that the energy-momentum tensor for a massive
($m$) spinning ($j$) pointlike source is \[
T_{00}=m\delta^{\left(2\right)}\left(\vec{r}-\vec{r}_{1}\right),\qquad T_{\phantom{i}0}^{i}=\frac{1}{2}j\epsilon^{ij}\partial_{j}\delta^{\left(2\right)}\left(\vec{r}-\vec{r}_{1}\right),\qquad T_{ij}=0.\]
 For two such conserved sources scattering amplitude was computed
in \cite{gullu} as\begin{eqnarray*}
4A & = & \int d^{3}x\,\left\{ -2T_{\mu\nu}^{\prime}\left[\beta\Box^{2}+\frac{1}{\kappa}\Box\right]^{-1}T^{\mu\nu}+T^{\prime}\left[\beta\Box^{2}+\frac{1}{\kappa}\Box\right]^{-1}T-T^{\prime}\left[\left(8\alpha+3\beta\right)\Box^{2}-\frac{1}{\kappa}\Box\right]^{-1}T\right\} .\end{eqnarray*}
 From the nonspinning case, the only added part will be\[
-4T_{i0}^{\prime}\left(\beta\Box^{2}+\frac{1}{\kappa}\Box\right)^{-1}T^{i0}=-\frac{j_{1}j_{2}}{\beta m_{g}^{2}}\partial_{i}\delta^{\left(2\right)}\left(\vec{r}-\vec{r}_{1}\right)\left(\frac{1}{\Box}-\frac{1}{\Box-m_{g}^{2}}\right)\partial_{i}\delta^{\left(2\right)}\left(\vec{r}-\vec{r}_{2}\right).\]
 After carrying out the space integrations, it reads \[
-4T_{i0}^{\prime}\left(\beta\Box^{2}+\frac{1}{\kappa}\Box\right)^{-1}T^{i0}=-\frac{j_{1}j_{2}}{2\pi\beta}\text{\ensuremath{K_{0}}}\left(m_{g}\left|\vec{r}_{1}-\vec{r}_{2}\right|\right),\]
 for \textbf{$\vec{r}_{1}\not=\vec{r}_{2}$}. Then, the total Newtonian
potential energy, $U=A/\text{time}$, becomes\textbf{\[
U=\frac{\kappa}{8\pi}\left(m_{1}m_{2}+4m_{g}^{2}j_{1}j_{2}\right)\text{\ensuremath{K_{0}}}\left(m_{g}\left|\vec{r}_{1}-\vec{r}_{2}\right|\right)-\frac{\kappa}{8\pi}m_{1}m_{2}\text{\ensuremath{K_{0}}}\left(m_{s}\left|\vec{r}_{1}-\vec{r}_{2}\right|\right).\]
 }Since $j_{1}$ and $j_{2}$ could be of any sign, the part coming
from the spin-spin interaction can be repulsive or attractive. In
the BHT limit the last term disappears.

\section*{Appendix B: The $\alpha$, $\beta$ theory with auxiliary fields}

Consider the quadratic Lagrangian (\ref{eq:Non-linear_action}) in
three dimensions. Using two auxiliary fields $\phi$ and $f_{\mu\nu}$,
one can rewrite it as\[
\mathcal{L}=\frac{1}{\kappa}\sqrt{-g}\left[R-f^{\mu\nu}G_{\mu\nu}-\phi R+\frac{m_{1}^{2}}{2}\phi^{2}+\frac{m_{2}^{2}}{4}\left(f^{\mu\nu}f_{\mu\nu}-f^{2}\right)\right],\]
 where $m_{1}^{2}=-\frac{4}{\kappa\left(8\alpha+3\beta\right)}$ and
$m_{2}^{2}=-\frac{1}{\kappa\beta}$. After linearization around flat
spacetime, we have\begin{align*}
\kappa\mathcal{L}_{linearized} & =-\left(\frac{1}{2}h^{\mu\nu}+f^{\mu\nu}\right)\mathcal{G}_{\mu\nu}^{L}-\phi R_{L}-\frac{2}{\kappa\left(8\alpha+3\beta\right)}\phi^{2}-\frac{1}{4\kappa\beta}\left(f^{\mu\nu}f_{\mu\nu}-f^{2}\right).\end{align*}
 For $8\alpha+3\beta=0$, $\phi$ decouples, and $f_{\mu\nu}$ can
be eliminated to yield the action describing spin-2 field with a Pauli-Fierz
mass\cite{bht}. But, for generic $\alpha$ and $\beta$, one has
to find a way to decouple $\phi$ , $f_{\mu\nu}$, and $h_{\mu\nu}$
keeping in mind that there should be a kinetic term for the $\phi$
field. This is possible by rescaling $h_{\mu\nu}$, but we have not
pursued this \cite{hindawi}.

\section*{Appendix C: Linearized field equations in the de Sitter background}

In the body of the text, we worked mostly at the level of the action.
To check our results at the level of the field equations, some of
the computations in this Appendix are needed. The trace of the linearized
field equation is \[
\left(8\alpha+3\beta\right)\Box R_{L}+\left[\frac{6\left(4\alpha+\beta\right)}{\ell^{2}}-a\right]R_{L}=0,\]
 where $\bar{g}^{\mu\nu}\mathcal{G}_{\mu\nu}^{L}=-\frac{R_{L}}{2}$
was used. Without further ado, let us list the results of somewhat
tedious, yet relevant computations:\begin{align*}
\Box\mathcal{G}_{00}^{L} & =\frac{t^{3}}{2\ell^{3}}\left[\left(\nabla^{2}\ddot{f}+\frac{5}{t}\nabla^{2}\dot{f}-\nabla^{2}\nabla^{2}f\right)-\frac{4}{t}\nabla^{2}p-\frac{3}{t^{2}}\nabla^{2}f-\frac{2\ell^{3}}{t^{5}}R_{L}\right],\end{align*}
 \[
\Box\mathcal{G}_{0i}^{L}=\frac{t^{3}}{2\ell^{3}}\partial_{i}\left(\ddot{p}+\frac{3}{t}\dot{p}-\nabla^{2}p-\frac{2}{t^{2}}p-\frac{2}{t}\nabla^{2}f\right)+\frac{t^{3}}{2\ell^{3}}\epsilon_{ij}\partial_{j}\left(\ddot{\sigma}+\frac{3}{t}\dot{\sigma}-\nabla^{2}\sigma-\frac{2}{t^{2}}\sigma\right),\]
 \begin{align*}
\Box\mathcal{G}_{ij}^{L} & =\frac{t^{3}}{2\ell^{3}}\left(\delta_{ij}+\hat{\partial}_{i}\hat{\partial}_{j}\right)\left(\ddot{q}+\frac{5}{t}\dot{q}+\frac{1}{t^{2}}q-\nabla^{2}q-\frac{2}{t^{2}}\nabla^{2}f\right)\\
 & \phantom{=}-\frac{t^{3}}{2\ell^{3}}\hat{\partial}_{i}\hat{\partial}_{j}\left(\overset{\dots}{p}+\frac{5}{t}\ddot{p}+\frac{1}{t^{2}}\dot{p}-\nabla^{2}\dot{p}-\frac{4}{t}\nabla^{2}p-\frac{2}{t^{2}}\nabla^{2}f\right)\\
 & \phantom{=}-\frac{t^{3}}{2\ell^{3}}\left(\epsilon_{ik}\hat{\partial}_{k}\hat{\partial}_{j}+\epsilon_{jk}\hat{\partial}_{k}\hat{\partial_{i}}\right)\left(\overset{\dots}{\sigma}+\frac{5}{t}\ddot{\sigma}+\frac{1}{t^{2}}\dot{\sigma}-\nabla^{2}\dot{\sigma}-\frac{2}{t}\nabla^{2}\sigma\right).\end{align*}
 $\mathcal{G}_{\mu\nu}^{L}$, and $R_{L}$, computed in the body of
the text, together with the Bianchi identity (\ref{eq:Bianchi_id}),
and the above results are sufficient to study the field equations.

\end{document}